# Computational complexity of spin-glass three-dimensional (3D) Ising model

Zhidong Zhang

Shenyang National Laboratory for Materials Science, Institute of Metal Research, Chinese Academy of Sciences, 72 Wenhua Road, Shenyang, 110016, P.R. China

Corresponding author's e-mail: zdzhang@imr.ac.cn

**Abstract:**

In this work, the computational complexity of a spin-glass three-dimensional (3D) Ising model (for the lattice size N = lmn, where l, m, n are the numbers of lattice points along three crystallographic directions) is studied. We prove that an absolute minimum core (AMC) model consisting of a spin-glass 2D Ising model interacting with its nearest neighboring plane, has its computational complexity $O(2^{mn})$. Any algorithms to make the model smaller (or simpler) than the AMC model will cut the basic element of the spin-glass 3D Ising model and lost many important information of the original model. Therefore, the computational complexity of the spin-glass 3D Ising model cannot be reduced to be less than $O(2^{mn})$ by any algorithms, which is in subexponential time, superpolynomial.

# 1. Introduction

In nature, order and disorder are two different states in a material. The study on the phase transition between the order and the disorder states is a very active topic in condensed matter physics and mathematical physics. In a ferromagnet, the magnetic order (ferromagnetic state) is emerged at the critical point, which is transformed from the disorder state (paramagnetic state) as temperature is decreased. A spin glass is a disordered magnet, where the spins of the component atoms are not aligned in a regular pattern [1-6]. The magnetic disorder of a spin glass compared to the magnetic order in a ferromagnet is similar to the positional disorder of a glass containing frustration compared to quartz. It is well-known that the atomic bond structure in a window glass or any amorphous solid is highly irregular, and this remains true also in the presence of frustration. In contrast, a crystal like quartz has a uniform pattern of atomic bonds. Analogously, the magnetic spins in a ferromagnet all align in the same direction in its ground state, while all the spins in a spin glass are frozen in a disorder ground state, aligning randomly to different directions. In a certain sense, the spin-glass state is an ordered state with disorder orientations of spins, in which the spins align disorderly in space, but may remain ordered (and/or unchanged) with the time evolution (associated to the onset of the spontaneous replica symmetry breaking). This indicates a fact that in a spin glass system the long-range correlation does not exist in space, but it may occur in the time evolution. Usually, the interactions between individual spins in a spin glass are a mixture of roughly equal numbers of ferromagnetic bonds (favoring the parallel alignment of neighboring spins) and antiferromagnetic bonds (favoring the antiparallel alignment of neighbors), and there may exist the bonds satisfying the frustration requirements on elementary plaquettes. This would cause the worst case for spin configurations and for computational complexity of the spin-glass model. The frustration states of spins are created in a spin glass, by the competitive actions of ferromagnetic and antiferromagnetic interactions, to form the patterns of aligned and misaligned spins, which, neglecting frustration, are similar to distortions in the geometry of atomic bonds in a glass. They may also create situations of degenerate, where more than one geometric arrangement of spins (or atoms) are stable in a spin glass (or a glass), since the presence of frustrated plaquettes makes the spin-glass case more tanglesome. The metastable state is termed to describe spin glasses and the complex internal structures that arise within them, because they are formed in some relatively stable configurations other than the lowest-energy configuration. It is fruitful to study the computational complexity of the structures of the spin glass, which is an important topic in computer science in addition to physics, chemistry, and materials science.

The Ising model is known to not only describe the transition from a paramagnetic to a ferromagnetic phase in a magnetic lattice, but also become a paradigm for several different systems, including antiferromagnets, lattice gases, and large biological molecules. The Ising model [7-15] is established based on the interaction between spins, when a spin coordinate $\sigma$ is associated with each lattice point of a crystal, which is considered as a scalar quantity and can achieve either of two values $\sigma = \pm 1$ (corresponding to either spin-up or spin down state). Usually, only the interaction energy between two spins located at the nearest neighbors of the lattice points is taken into account. The interaction may cause an ordered ground state (such as ferromagnetic, antiferromagnetic, ferrimagnetic, etc.) and also the order-disorder phase transitions upon competition with thermal activity. The thermodynamic and magnetic properties of the Ising model which contains N lattice points can be determined from the partition function. For a one-dimensional (1D) Ising model [7], spins are located at each lattice point of a chain with m sites, and Ising himself proved that there is no phase transition at finite temperature. For a ferromagnetic two-dimensional (2D) Ising model [8,12], Onsager obtained the exact solution of a rectangular lattice, which illustrates the emergence of singularity of specific heat at/near the critical point of the ferromagnetic-to-paramagnetic phase transition, which comes from a non-singularity function (Hamiltonian). In the 2D Ising model Onsager and Kaufman studied [8,12], the rectangular lattice is constructed by m rows and n columns (with $N = mn$ sites). The Onsager-Kaufman procedure can be extended to be appropriate for other 2D Ising models on triangle or hexagonal lattice, but not for those with crossing interactions. For a ferromagnetic three-dimensional (3D) Ising model [9-15], as it is constructed by l planes, in each plane there are m rows and n columns (in total with $N = lmn$ sites), the situation becomes very complicated, since not only the number of lattice points increases greatly, but also the nature of three dimensions causes many complicities, such as non-trivial topological structure, non-locality, long-range entanglement (even only the nearest neighboring interactions are considered). On the observation of the formula of the partition functions of a ferromagnetic 3D Ising model, the author conjectured that the non-trivial topological structures of the ferromagnetic 3D Ising model can be trivialized in a higher dimensional space and the ferromagnetic 3D Ising model can be realized as the free statistic model on (3+1) dimensions with weight factors (i.e., topological/geometrical phases) on eigenvectors [13]. Zhang then studied the mathematical structure of the ferromagnetic 3D Ising model [14]. Recently, Zhang, Suzuki and March developed a Clifford algebra approach for the ferromagnetic 3D Ising model [15], which gives a positive answer to the Zhang's two conjectures [13]. However, the interest of the present work will focus on a much more complicated system, i.e., a spin-glass 3D Ising model with randomly distributed positive and negative interactions between spins, and containing frustration, while its limit case with all interactions positive and the frustration switched off is the random ferromagnetic 3D Ising model. The purpose of this work is not to find the exact solution of the spin-glass 3D Ising model, but only to study its computational complexity.

The Hamiltonian for a spin-glass 3D Ising model is given by [1-6]

$$H = -\sum_{<i,j>} J_{ij} S_i S_j$$

Here we have spins arranged on a 3D lattice with only nearest neighboring interactions similar to the ferromagnetic Ising model, but now also frustration is present. Our physical model is a simple orthorhombic lattice with m rows and n sites per row in one of l planes, in total the number of lattice sites N = lmn. The main difference with the ferromagnetic Ising model is that the interaction $J_{ij}$ with different signs is randomly distributed in the present spin-glass Ising model, and also frustration is present. As usual, the probability of finding the spin-glass 3D Ising lattice in a given configuration, and a fixed replica, at the temperature T, is proportional to exp{-E$_c$/k$_B$T}, where E$_c$ is the total energy of the configuration and k$_B$ is the Boltzmann constant. The thermodynamic functions for the spin-glass 3D Ising model can be found from knowledge of the partition function Z, after mediating ln Z over disorder. The partition function Z for the spin-glass 3D Ising lattice can be expressed in a fixed replica as [8,12,13]:

$$Z = \sum_{\substack{all \\ configurations}} e^{n_c K_{ij} + n'_c K'_{ij} + n''_c K''_{ij}}$$

Here $n_c$, $n'_c$ and $n''_c$ are integers depending on the configuration of the lattice. We introduce variables $K_{ij} \equiv J_{ij}/k_B T$, $K'_{ij} \equiv J'_{ij}/k_B T$ and $K''_{ij} \equiv J''_{ij}/k_B T$, instead of $J_{ij}$, $J'_{ij}$ and $J''_{ij}$, for the randomly distributed interactions along three crystallographic directions of the lattice. Similar to the ferromagnetic case, the partition function of the spin-glass 3D Ising lattice in a fixed replica may be written in forms of three transfer matrices in forms of direct products of matrices [9,13-15]. However, because of the randomness of interactions, the differences with the ferromagnetic case are: 1) It is impossible to apply the periodic condition to reduce the size of the transfer matrices and to obtain the eigenvalues in forms of $\lambda_i^m$. 2) The product over j in the transfer matrices cover from 1 to lmn. 3) It is not possible to define a unique K*. Nevertheless, the non-locality as well as the non-trivial topological structure exist also in the spin-glass 3D Ising system, similar to the ferromagnetic 3D Ising lattice [13-15], caused by the interactions along the third dimension in the 3D space.

The Edwards–Anderson model [2], in which only interactions between the nearest neighboring spins are considered, can be solved via replica trick for the critical temperature and a spin glass phase is observed to exist at low temperatures [16]. In order to determine the partition function for this system, one needs to average the free energy over all possible values of interactions $J_{ij}$. The distribution $P(J_{ij})$ of values of $J_{ij}$ is taken to be a Gaussian with a mean $J_0$ (taken zero in [2]) and a variance $J^2$. Solving for the free energy using the replica method, below a certain temperature, it was found that the spin glass phase of the system exists which is characterized by a vanishing magnetization along with a non-vanishing value of the two point correlation function between spins at the same lattice point but at two different replicas: $q = <S_i^\alpha S_i^\beta>|_{\alpha \neq \beta} \neq 0$, where $\alpha$ and $\beta$ are replica indices. q is the order parameter for the paramagnet to spin glass phase transition. Hence the new set of order parameters describing the possible phases consists of both m and q (however m = 0 in [2]). In this work, we will focus on the Edwards–Anderson model supplemented also with non-zero mean $J_0 \neq 0$. However, we are interested only on its exact solution in the worst case for its complexity, which is not like the previous work in which usually it is treated by the approximation averaging the free energy over all possible values of interactions or by the replica method in a mean-field type of approximation.

The model of Sherrington and Kirkpatrick is important, considered by the authors to be an exactly solvable model of a spin glass [3,5], which is an Ising model with long-range ferromagnetic as well as antiferromagnetic couplings for frustrated states. However, it corresponds to a mean-field approximation of spin glasses describing the slow dynamics of the magnetization and the complex non-ergodic equilibrium state. The equilibrium solution of the Sherrington-Kirkpatrick model, being thermodynamically misleading and incorrect in the T to zero limit, has been improved by Parisi within the non-symmetric replica [4]. It was revealed that the complex nature of the spin glass low-temperature phase is characterized by ergodicity breaking, ultrametricity and non-selfaverageness [17-19]. Further developments led to the creation of the cavity method, which allowed study of the low-temperature phase without replicas. A rigorous proof of the Parisi solution has been provided in [20,21]. In this work, we are interested in a spin-glass 3D Ising model with the nearest neighboring interactions only. In this sense, we will exclude the Sherrington-Kirkpatrick model from our study, because it is with long range interactions, and with a mean-field approximation.

In what follows, we list the characters of the spin-glass 3D Ising model

**1) Topological effect:** The existence of topological effect in the ferromagnetic 3D Ising model has been pointed out in [11], which should be true for the spin-glass 3D Ising model. The combinatorial method of counting the closed graph for the 2D Ising model cannot be generated in any obvious way to the 3D problem (see page 366 in [11]). For the 3D Ising model, one encounters polygons with knots (see page 367 in [11]). The peculiar topological property is that a polygon in three dimensions does not divide the space into an "inside and outside" (see page 367 in [11]). The non-trivial topological structures in the ferromagnetic 3D Ising model are observed in [13-15].

**2) Randomness:** One of the most important characters of the spin-glass

systems is the randomly distributed interactions between spins, and the presence of frustration [1-6,22-24]. The randomness of the distribution of the interactions causes the computational complexity becomes much more complicated than that of the ferromagnetic counterpart. This is because the randomly distributed interactions may result in not only the random distribution of spin alignments, but also the frustrated plaquettes. Note that in a limit case if all the interactions are ferromagnetic and randomly distributed, a random ferromagnetic state without frustration may occur. Furthermore, the random interactions break down the translational symmetry, which usually exists in a crystal or a ferromagnet. This symmetrical breaking down leads to the invalidity of some approaches for a ferromagnetic system, such as the periodic boundary condition, a unique rotation angle for a local transformation if applied for the whole system, etc.

**3) Frustration:** Normally, frustration consists of geometrical frustration and compositional frustration. Geometrical frustration is an important feature in magnetism, where it stems from the topological arrangement of spins. A spin at a particular lattice (such as triangular, honeycomb, Kagome, etc.) can be frustrated because its two possible orientations, up and down, give the same energy, which is acted by its neighboring spins with antiferromagnetic couplings [1-6,25-28]. The ground state is multi-fold degenerate. The Ising model on a triangular lattice with nearest-neighboring spins coupled antiferromagnetically was studied in 1950 by Wannier [29]. A renewed interest in such spin systems with frustrated or competing interactions arose about two decades later, in the context of spin glasses and spatially modulated magnetic superstructures [30,31]. In spin glasses, frustration is augmented by stochastic disorder in the interactions. The compositional frustration is caused by competition of different bonds between different atoms such to produce frustrated plaquettes. In our spin-glass Ising model with simple orthorhombic lattice, the frustration is mainly resulted from the competition of different exchange couplings between spins, which is more like compositional frustration, and the rapid quenching of the system leads to frustrated plaquettes. This is because the frustration can occur owing to the competition between ferromagnetic and antiferromagnetic interactions between spins at a lattice (even without the particular requires for lattice geometry, for instance, a square lattice or a cubic lattice). The frustration occurs when the product of signs of all the interactions along a plaquette is negative.

4) **Non-ergodic behavior:** A so-called non-ergodic behavior happens in spin glasses below the freezing temperature $T_f$, since the system cannot escape from the ultradeep minima of the hierarchically-disordered energy landscape [6,32]. The hierarchical disorder of the energy landscape may be verbally characterized by a scenario that there are (random) valleys within still deeper (random) valleys within still deeper (random) valleys, ..., etc.

Finding the ground state of the spin-glass Ising model can be done by computing $H(\sigma)$ for accounting the combinatorial complexity for all $2^N$ possible $\sigma$'s [1,3,10,18,22,24]. The upper bound of the computational complexity of a spin-glass 3D Ising model is $O(2^N)$. It will be of interest to find the lower bound of the computational complexity of a spin-glass 3D Ising model, which will be done in section 2. However, the Zhang-Suzuki-March procedure [15] for the ferromagnetic 3D Ising model cannot be applicable for computing analytically the spin-glass 3D Ising model. This statement can be clearly seen from the following analysis. In the Onsager-Kaufman procedure [8,12] for the ferromagnetic 2D Ising model and the Zhang-Suzuki-March procedure [15] for the ferromagnetic 3D Ising model, a periodic boundary condition is employed along one of the three crystallographic directions (say, along the x direction). For purpose of the symmetry, it is assumed that the m-th row in each plane of the crystal interacts with the first row in that plane. To do so, we actually apply the cylindrical crystal model preferred by Onsager [12] and Kaufman [8] for the ferromagnetic 2D Ising model, and Zhang [13,14] for the ferromagnetic3D Ising model, in which we wrap our crystal on cylinders. This causes the simplicity of the calculations, since the dominant contribution to the partition function is the largest eigenvalue of the models [8,12-15]. The randomness of interactions between spins in this spin-glass 3D Ising model breaks down the periodic boundary condition. Thus, the spin-glass 3D Ising model cannot be solved by using the periodic boundary condition. Furthermore, the Largest Eigenvalue Principle used in Zhang-Suzuki-March procedure [15] cannot be utilized for computing analytically the spin-glass 3D Ising model. Also because of the random distribution of interactions, the rotation angles for local transformation in the spin-glass 3D Ising model are randomly distributed, which causes great computational complexity. Therefore, its computational complexity cannot be reduced by the periodic boundary condition, the Largest Eigenvalue Principle and local transformation in the Zhang-Suzuki-March procedure [15]. The only thing in our previous work, which we use in this work, is a fact that there exists the long-range entanglement, due to the internal factors in the transfer matrices (as clearly seen in Eq. (15) in [14], Eq. (A10) in [15], Eq. (3) in [33] and Eq. (73) in [9]), which is the character of the 3D Ising model.

In this work, we shall be interested only in the spin-glass 3D Ising lattice with strongly competing interactions in general cases (or worst cases for computational complexity), in which, in the presence of frustration, neither ferromagnetic nor antiferromagnetic interaction is dominant so that it is not easy to figure out the ground state of the system to be ferromagnetic, antiferromagnetic or spin glass.

## 2. Computational complexity of spin-glass 3D Ising model

In this section, we prove four theorems for the computational complexity of the spin-glass 3D Ising model:

**Theorem 1** The core model of the spin-glass 3D Ising model consists is much more complex than an absolute minimum core (AMC) model consisting of a spin-glass 2D Ising model interacting with its nearest neighboring plane.

Proof:

In the ferromagnetic 3D Ising model, it was revealed in Eq. (15) in [14], Eq. (A10) in [15], Eq. (3) in [33] and Eq. (73) in [9] that many internal factors exist in the transfer matrices, which show the non-local behaviors of the spins. It was uncovered [14,15] that the interaction between the most neighboring spins along the third dimension behaves as an interaction between two spins located far from each other, via mn spins in the plane. The same effect happens for every interaction along the third dimension in the ferromagnetic 3D Ising model, Although the Ising model looks like to be fully locally defined in the original Ising spin variable language, the set of all allowed states contribute to partition function and free energy in a way of all spins entangled. The non-locality shows up in the alternative Clifford algebra description, defined through auxiliary fermionic $\Gamma$-operators. The non-locality exists not only in the space of description of $\Gamma$-operators, but also in the space of all the Ising spin states. Although it is not evidently and clearly seen in the latter space, the descriptions in the two different spaces are connected by a series of equalities, so the same non-local effect appears in both the spaces. The same is true for the spin-glass 3D Ising model. This is because the internal factors are the intrinsic property of a 3D Ising model, due to its topology, which does exist naturally also in the spin-glass 3D Ising model. The existence of randomness in the spin-glass 3D Ising model does not change this character. Therefore, it is clear that there exists an absolute minimum core (AMC) model in the spin-glass 3D Ising model, in which the entanglements between the spins should not be broken. The AMC model of the spin-glass 3D Ising model consists of a spin-glass 2D Ising model interacting with its nearest neighboring plane. In the AMC model, a plaquette within the two neighboring planes may show frustration if the condition for frustration along the plaquette is satisfied. This AMC model represents the intrinsic characters of the spin-glass 3D Ising model: nonplanarity graphs, long-range entanglement, non-locality, frustrations, etc. In addition, it is big enough to illustrate randomness of interactions. Because of such a complicated AMC model of the spin-glass 3D Ising model, its computational complexity cannot be reduced further. The study on a model smaller (or simpler) than the AMC model will cut the AMC model and lost many important information of the original model. However, the core model of the spin-glass 3D Ising model is much more complex than the AMC model. This is because in some replicas, frustration in the 3D case could appear on closed polygons, which are higher than a plaquette. Such closed polygons cannot be included always in two neighboring planes. Indeed, more than two neighboring planes must be considered, if we consider all the possible frustrations in the 3D lattice. This makes that the computational complexity of the core model of the spin-glass 3D Ising model is much higher than that of the AMC model.

∎

**Theorem 2** The computational complexity of a spin-glass 3D Ising model cannot be reduced to be less than l times the computational complexity of its absolute minimum core (AMC) model.

Proof:

The probability of all the states in the spin-glass 3D Ising model can be written as the direct product of the probabilities of the states respectively in all the sub-models with smaller sizes for independent computations. According to Theorem 1, the AMC model of the spin-glass 3D Ising model is constructed by a spin glass 2D Ising model interacting with one of its nearest neighboring planes. The spin-glass 3D Ising model consists of l terms of its AMC model, being a sub-model. The AMC model is the smallest sub-model that can represent all the intrinsic characters of the spin-glass 3D

Ising model. Any algorithms to solve exactly the spin-glass 3D Ising model must deal with at least one AMC model and the existence of any sub-model bigger than it will result in a larger computational complexity increasing exponentially with the size of the biggest sub-model. Because the lower bound of the computational complexity of the spin-glass 3D Ising model is determined by the size of the largest sub-model for independent computations, it should be as small as possible and reduced to be the AMC model. The AMC model follows the spin arrangements in a sequence of constructing first a plane, then stacking another plane, and one by one, for the construction of the 3D model. This is a simplest procedure for constructing a 3D model with the smallest computational complexity. Any other methods and/or algorithms to deal with a spin-glass 3D Ising model are more complex than the method to deal with l terms of the AMC model. This is because the AMC model is the simplest arrangement of spins with natural sequence, while maintaining the characters of the 3D model. Other spin arrangements will introduce more complexities, and face to deal with at least one sub-model bigger than the AMC model. For instance, if one chose the first site of the second plane as the fourth site, and so on, after one covered the first two planes, starting from the third plane, the long-range entanglements caused by many-body interactions would appear even with more disorder and without regular pattern. One would face how to represent the transfer matrices in more complicated ordering in terms of the direct products of Pauli matrices, the problem would become much more complicated by such a method (and any other algorithms). Therefore, the direct product of the probabilities of the states in l terms of the AMC models is the lower bound for accounting the probability of all the states in the spin-glass 3D Ising model.

∎

Therefore, any algorithms, which use any approximations and/or break our AMC model, cannot result in the exact solution of the spin-glass 3D Ising model. The computational complexity of the AMC model is the lower bound for computation of the spin-glass 3D Ising model.

**Theorem 3** The computational complexity of the AMC model of a spin-glass 3D Ising model cannot be reduced to be less than $O(2^{mn})$ by any algorithms. The computational complexity in the order of $O(2^{mn})$ is much less than $O(2^N)$, but much larger than and cannot be reduced to polynomial time $O(N^p)$.

Proof:

For the spin-glass 3D Ising model with m rows, n lines and l planes, the total number of the sites in the lattice is $N = lmn$. The number of the lattice points in an AMC model of the spin-glass 3D Ising model (a plane) is $M = mn$. The computational complexity of the AMC model of a spin-glass 3D Ising model cannot be reduced to be less than $O(2^{mn})$ by any algorithms, because of the non-locality in the system. Indeed, one has to take into account all the combinatorial complexity of this AMC model.

First, we compare $O(2^{mn})$ with $O(2^N)$, where $N = lmn$ with $l \to \infty$, $m \to \infty$, $n \to \infty$ in the thermodynamic limit. We have: $\lim\limits_{\substack{l \to \infty \\ m \to \infty \\ n \to \infty}} \dfrac{2^N}{2^{mn}} = \lim\limits_{l \to \infty} 2^l \to \infty$ The computational complexity in the order of $O(2^{mn})$ is much less than $O(2^N)$.

If we assume that $n = m = l$, we will have $n = m = l = N^{1/3}$ and $mn = N^{2/3}$. Suppose

$z^N = 2^{N^{2/3}}$, we will have $z = \left[2^{N^{2/3}}\right]^{\frac{1}{N}} = 2^{N^{-1/3}}$. As N → ∞, we have z → 1. However, since z is always larger than 1 (although it is approaching 1 as N → ∞), we can write it as (1 + ε) with ε → 0. Note that here ε ≠ 1/N. Clearly, $2^{mn}$ is equal to $(1 + ε)^N$, which is subexponential. We can prove that $O(2^{mn})$ is superpolynomial time (see below).

Next, we compare $O(2^{mn})$ with $O(N^p)$, where p is a finite number,

$$\lim_{\substack{l \to \infty \\ m \to \infty \\ n \to \infty}} \frac{(lmn)^p}{2^{mn}} = \lim_{\substack{m \to \infty \\ n \to \infty}} \frac{(mn)^{3p/2}}{2^{mn}} = \lim_{M \to \infty} \frac{M^{3p/2}}{2^M} = \lim_{M \to \infty} \frac{\left(M^{3p/2}\right)}{\left(2^M\right)} = \lim_{M \to \infty} \frac{\frac{3p}{2} M^{\left(\frac{3p}{2}-1\right)}}{2^M \ln 2}$$

$$= ... = \lim_{M \to \infty} \frac{\left(\frac{3p}{2}\right)!}{2^M (\ln 2)^{\frac{3p}{2}}} \to 0$$

In above, we used the L'Hôpital's rule and we assume that $l = (mn)^{1/2}$ without loss of generality, since l has the same order as m or n. Clearly, the computational complexity in the order of $O(2^{mn})$ is much larger than and cannot be reduced to polynomial time $O(N^p)$. That is subexponential, but superpolynomial.

∎

**Theorem 4** The computational complexity of a spin-glass 3D Ising model cannot be reduced to be less than $O(2^{mn})$ by any algorithms, which is subexponential and superpolynomial.

Proof:

It is valid as an immediate consequence of Theorems 1-3.

∎

The results obtained in this work provide a proof of Theorems on NP-complete problems [34,35].

**Acknowledgements**

This work has been supported by the National Natural Science Foundation of China under grant numbers 51590883 and 51331006, by the State Key Project of Research and Development of China (No.2017YFA0206302). The author is grateful to Fei Yang for understanding, encouragement, support and discussion.